# Nuclide Production Cross Sections for $^{59}$Co and $^{nat}$Cu Irradiated with 0.2 GeV and 2.6 GeV Protons and 0.2 GeV/Nucleon Carbon Ions

Yu.E. Titarenko, V.F. Batyaev, V.M. Zhivun, A.B. Koldobsky, Yu.V. Trebukhovsky, E.I. Karpikhin, R.D. Mulambetov, S.V. Mulambetova, Yu.V. Nekrasov, A.Yu. Titarenko, K.A. Lipatov, B.Yu. Sharkov, A.A. Golubev, A.D Fertman, V.I Turtikov, A.V. Kantsyrev, I.V Roudskoy, G.N. Smirnov[1], V. S. Barashenkov[2], K.K. Gudima[3], M.I. Baznat[3], S.G. Mashnik[4], R.E. Prael[4]

[1] *Institute for Theoretical and Experimental Physics (ITEP), B.Cheremushkinskaya 25, 117259 Moscow, Russia*
[2] *Joint Institute for Nuclear Research (JINR), 141980, Dubna, Moscow Reg., Russia*
[3] *Institute of Applied Physics Academy of Science of Moldova, Chisinau, Moldova*
[4] *Los Alamos National Laboratory (LANL), Los Alamos, NM 87545, USA*

**Abstract** –*The results of experimental determining the cross sections for residual nuclide production in interactions of 200 MeV/A $^{12}$C ions and of 200 MeV and 2600 MeV protons with thin $^{Nat}$Cu, $^{59}$Co, and $^{27}$Al samples are presented. The residual products are measured by g-spectrometry with Ge detector of 1.8 keV resolution in the 1332 keV $^{60}$Co g-line. The measurement data are compared with the simulation results obtained by LANL and JINR codes.*

## I. INTRODUCTION

The recent years have seen an extensive development of ion accelerators, which necessitates prompt designing of their radiation shielding.

With the new acceleration-accumulation facilities under design and construction in Germany (SIS-200), Russia (TWA-ITEP), and Japan (MUSES-RIKEN) and pending the heavy-ion accelerators to resolve the problems of inertial thermonuclear fusion, the radiation and ecological safety issues are becoming particularly important in view of the beam intensity increase up to $10^{13}$-$10^{15}$ particle/pulse aimed at by the said projects. [1-4]

The present-day lack of any real data for estimating particle loss in acceleration channels, together with the lacking necessary nuclear cross sections, makes it difficult to reach the required accuracy of estimating the activation of the accelerator structure materials, given such high beam intensities. Therefore, determination of the cross sections for residual nuclide production in experimental samples irradiated by different-energy multiply charged ions has become a challenge to those who try to obtain the required nuclear data directly and to use the experimental data in verifying the simulation codes used to accumulate the data. Besides, comparing our obtained "ion" cross sections for residual nuclide production with the "proton" cross sections obtained for experimental samples of identical compositions at close kinetic energies is very important with a view to validating and updating the existing nuclear models.

## II. TECHNIQUES FOR DETERMINING THE PRODUCTION CROSS-SECTIONS OF PRODUCT NUCLEI

The $^{27}$Al(p,x)$^{7}$Be, $^{27}$Al(p,x)$^{22}$Na, and $^{27}$Al(p,x)$^{24}$Na monitor reaction-based techniques for determining the production cross-sections of radioactive product nuclei (spallation products, SP) in the proton reactions have been described in sufficient detail in. [5-7] In the case of ions, however, the techniques have been modified because of lacking any reliable monitor reaction data.

By analogy with independent and cumulative cross-sections for SP production, we introduce the concept of independent and cumulative reaction rates:

$$R^{ind}=\sigma^{ind}(E)\cdot\Phi(E),\ \text{and}\ R^{cum}=\sigma^{cum}(E)\ \Phi(E) \qquad (1)$$

where $R^{ind}$ and $R^{cum}$ are, respectively, the independent and cumulative rates of a nuclide production, $s^{-1}$; $\sigma^{ind}(E)$ and $\sigma^{cum}(E)$ are, respectively, the independent and cumulative cross sections of a nuclide production, barn; Φ is the flux density of the projectiles (protons or ions), $cm^2\cdot s^{-1}$.

Using, then, the approach described in [5-7], we can calculate the independent and cumulative reaction rates as

$$R_1^{cum}=\frac{A_0}{N_T\cdot h_1\cdot e_1}\cdot\frac{1}{F_1'} \qquad (2).$$

$$R_1^{cum} = \frac{A_1}{N_T \cdot \eta_2 \cdot \varepsilon_2 \cdot \nu_1} \cdot \frac{\lambda_2 - \lambda_1}{\lambda_2} \cdot \frac{1}{F_1'} \qquad (3),$$

$$R_2^{ind} = \left( \frac{A_2}{F_2'} + \frac{A_1}{F_1'} \cdot \frac{\lambda_1}{\lambda_2} \right) \cdot \frac{1}{N_T \cdot \eta_2 \cdot \varepsilon_2} \qquad (4),$$

$$R_2^{cum} = R_2^{ind} + \nu_1 \cdot R_1^{cum} =$$

$$= \left( \frac{A_1}{F_1'} + \frac{A_2}{F_2'} \right) \cdot \frac{1}{N_T \cdot \eta_2 \cdot \varepsilon_2} \qquad (5).$$

where $A_0$, $A_1$, and $A_2$ are the factors determined by the least-squares fitting of the experimental count rates of nuclides *1* and *2*; $N_T$ is the number of nuclides in an irradiated experimental sample; $\eta_1$, and $\eta_2$ are the absolute quantum yields of nuclides *1* and *2* at energies $E_1$ and $E_2$; $\varepsilon_1$ and $\varepsilon$ are the absolute spectrometer effectivenesses at $E_1$ and $E_2$; $\nu_1$ is the branching ratio; $F_1'$ and $F_2'$ are saturation functions of nuclides *1* and *2*.

The errors of the measured reaction rates were calculated by the standard error transfer formulas.

Having determined the SP production rates, we can calculate the respective cross sections by formulas (1). This approach to calculating the SP production cross sections makes it possible to use the particle flux density value obtained using monitor reactions or current transformer.

## III. IRRADIATION OF EXPERIMENTAL SAMPLES

The target irradiations by the external proton beams from the ITEP U-10 accelerator have been described in [7]. The experiments were made using Cu samples high-enriched with $^{63}$Cu (99.6%) and with $^{65}$Cu 98.7%) and the high chemically pure Co and Al samples (99.9%).

In the ion experiments, use was made of the external ion beam from the modified ITEP U-10 – TWA accelerator facility, which has been operative since 2002 [2]. The facility combines the actual U-10 proton synchrotron the newly-constructed second heavy-ion ring. The medium-mass He-like ions ($^{12}$C ions alone can be irradiated now) are accelerated in fore injector up to 1.6 MeV/nucleon and are then injected to the new booster ring. Having been accelerated there up to 700 MeV/nucleon, the 250-ns duration ion beam are injected to the old synchrotron ring due to changing charge from He-like ion to nucleus when moving through the electron-baring foil. Fig. 1 is a schematic of ion beam irradiation of a target.

Contrary to the proton beam experiments, the SP production cross sections for $^{nat}$Cu , $^{59}$Co, and $^{27}$Al in the ion beam were determined by irradiating the sandwich samples of six 10.5-mm diameter foils (Cu+Cu – Co+Co – Al+Al). In each of the foil pairs, only the second foils (of 0.25 mm, 0.40 mm, and 0.10 mm thickness, respectively) were spectrometered. In such a manner, the systematic errors arising from recoil nuclide escape from the samples have been avoided.

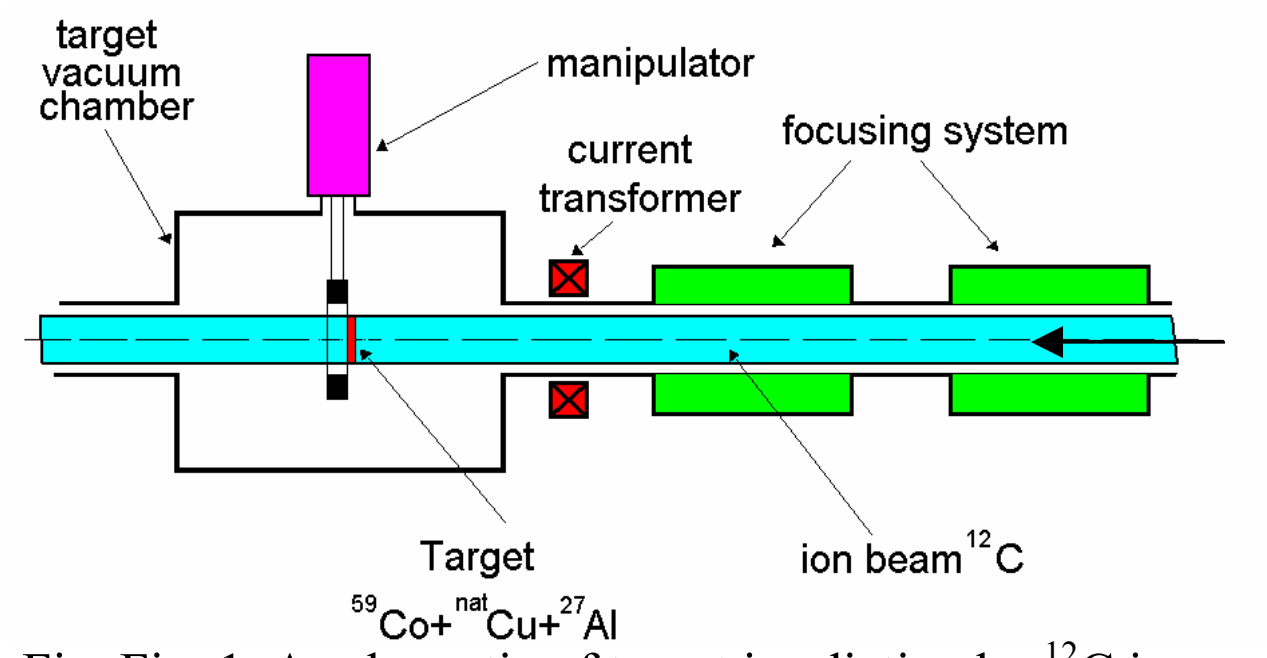

Fig. Fig. 1. A schematic of target irradiation by $^{12}$C ion beam.

## IV. MONITORING OF THE PROTON AND ION BEAMS

In the proton experiments, the 0.2 GeV and 2.6 GeV beams were monitored by the activation technique described in.[5-7]

In the ion experiments, the BERGOZ FCT-082-20:1 current transformer was used to monitor the beam. The FCT output signals were recorded by the TDS-220 digital oscilloscope of 2 ns time resolution. Fig. 2 is the oscillogram of the ion beam pulse fine structure.

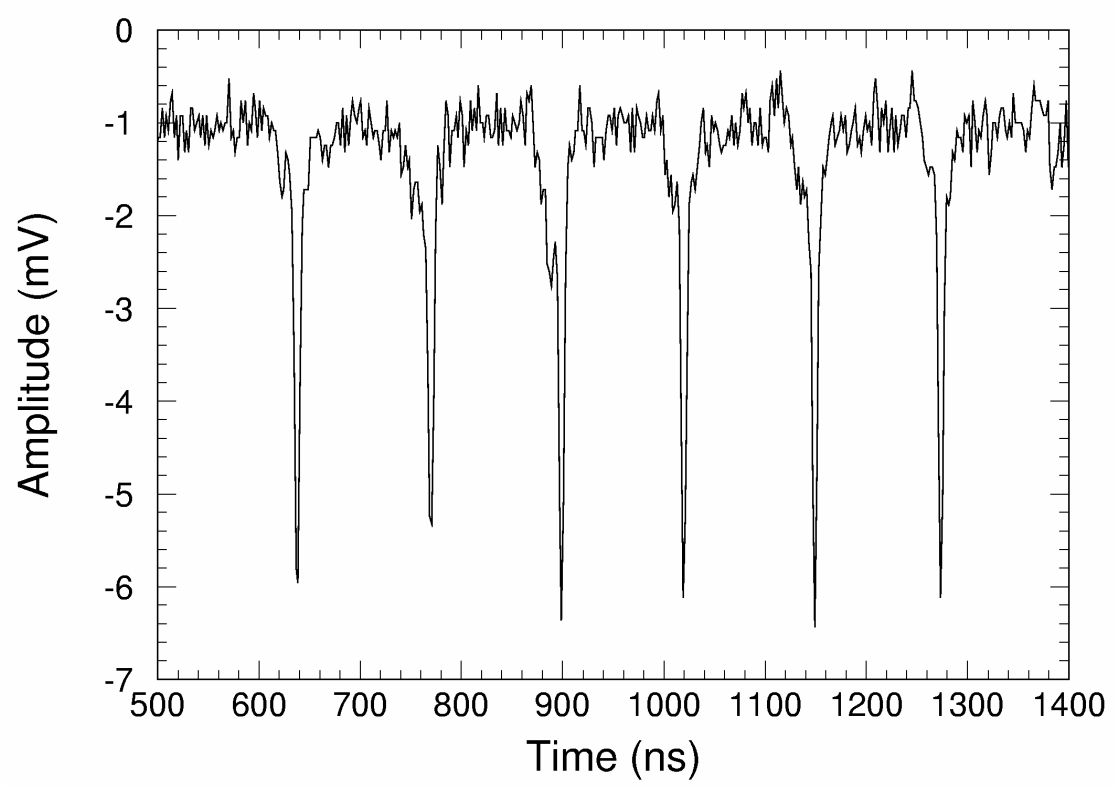

Fig. 2. The timebase oscillogram of the 200 MeV/nucleon $^{12}$C ion beam pulse from the ITEP U-10 – TWA facility.

With the ion beam intensity attained (~$3\cdot10^7$ cm$^{-2}$s$^{-1}$), the irradiation time was 61200 s at a 0.201 s$^{-1}$ pulse repetition rate. The $^{12}$C ion flux density was found to be $(3.34\pm0.24)\cdot10^7$ cm$^{-2}\cdot$s$^{-1}$ using the formula discussed in [8]:

$$\Phi = \frac{I \cdot t \cdot k_{flux}}{K \cdot z \cdot e \cdot T_{irr} \cdot S_{sample}} \qquad (6),$$

where $I$ is the channel-by-channel sum in the oscillogram pulses that corresponds to the ion beam timebase (see Fig. 2) with due allowance for the constant component, V; $\Delta t$ is the time width of the oscilloscope channel (digitizing), s ; $k_{flux}$ is the ratio of ion number through experimental sample to the total beam ion number; $K$ is the current transformer signal transformation coefficient, V/A; $z$ is charge number of accelerated ions; $Q$ is elementary charge, Coul; $T_{irr}$ is irradiation time, s; $S_{sample}$ is cross section area of experimental sample, $cm^2$**.**

## V. EXPERIMENTAL RESULTS

Tables 1-3 present the results of measuring the cross sections for SP production in interactions of the 200 MeV/A $^{12}C$ ions with thin $^{nat}Cu$, $^{59}Co$, and $^{27}Al$ samples. As to the tentative ion results, they have been published.[9] Forty-three cross sections have been determined for $^{nat}Cu$, thirty seven cross sections for $^{59}Co$, and three cross sections for $^{27}Al$. For comparison, Tables 1-3 present also the SP production cross sections in the 0.2 and 2.6 GeV proton interactions with the same samples.[7] The Cu data obtained with high-enriched samples[7] were scaled to $^{nat}Cu$ making allowance for the $^{63,65}Cu$ content in $^{nat}Cu$.

The experimental cross sections for SP production in 200 MeV/A $^{12}C$ ion interactions with thin $^{nat}Cu$ and $^{59}Co$ samples were simulated by the LAQGSM and CASCADE codes. Fig. 3 shows the results of nuclide-by-nuclide comparison between the experimental and calculated SP production cross sections in the 0.2 GeV $^{12}C$ ion irradiation of the $^{nat}Cu$ and $^{59}Co$ samples. The experiment-calculation comparison method described in [5] was used to calculate the mean squared deviation factors $<F>$ that characterize the predictive power of the codes. The $<F>$ values have been found to be 1.77 and 2.77 for LAQGSM and CASCADE, respectively.

Tables 1-3 demonstrate that the lists of the SPs identified in the ion and proton interactions are actually the same. Comparing between the proton and ion cross sections shows that, given the absolute kinetic energy values, proportionality holds between the proton and ion cross sections (see Figs 4 and 5). Fig. 6 shows the statistics of all ratios of the SP cross sections in 2.4 GeV (0.2 GeV/A) $^{12}C$ ion irradiation of $^{59}Co$ and $^{nat}Cu$ to the SP cross sections in the 2.6 GeV proton irradiation of the same samples.

If the analysis of the proton and ion ($^{12}C$) irradiation of $^{nat}Cu$ and $^{59}Co$ is made disregarding the cross sections for production of SPs of the least masses and masses close to the target nucleus mass, then the cross section ratio will be a simple relation:

$$\left(\frac{s^{SP}_{^{12}C}}{s^{SP}_{p}}\right)_{mean} = 2.3\pm0.2 \text{ (for } ^{nat}Cu,\ ^{59}Co) \qquad (7)$$

At a ~2.6 GeV kinetic energy, this ion-to-proton cross section ratio is in a satisfactory agreement with the same ratio inferred from formulas based on an approximate geometric representation of the ion and proton interactions, which is

$$\frac{s^{SP}_{^{12}C}}{s^{SP}_{p}} = \frac{(A^{1/3}_{Tag} + 2\cdot A^{1/3}_{C})^2}{(A^{1/3}_{Tag} + 2\cdot A^{1/3}_{p})^2}\ , \qquad (8)$$

and equals 2.0 for $^{nat}Cu$ and 2.1 for $^{59}Co$.

This work was partially supported by the U.S. Department of Energy and by the Moldovan-U.S. Bilateral Grants Program, CRDF Project MP2-3025.

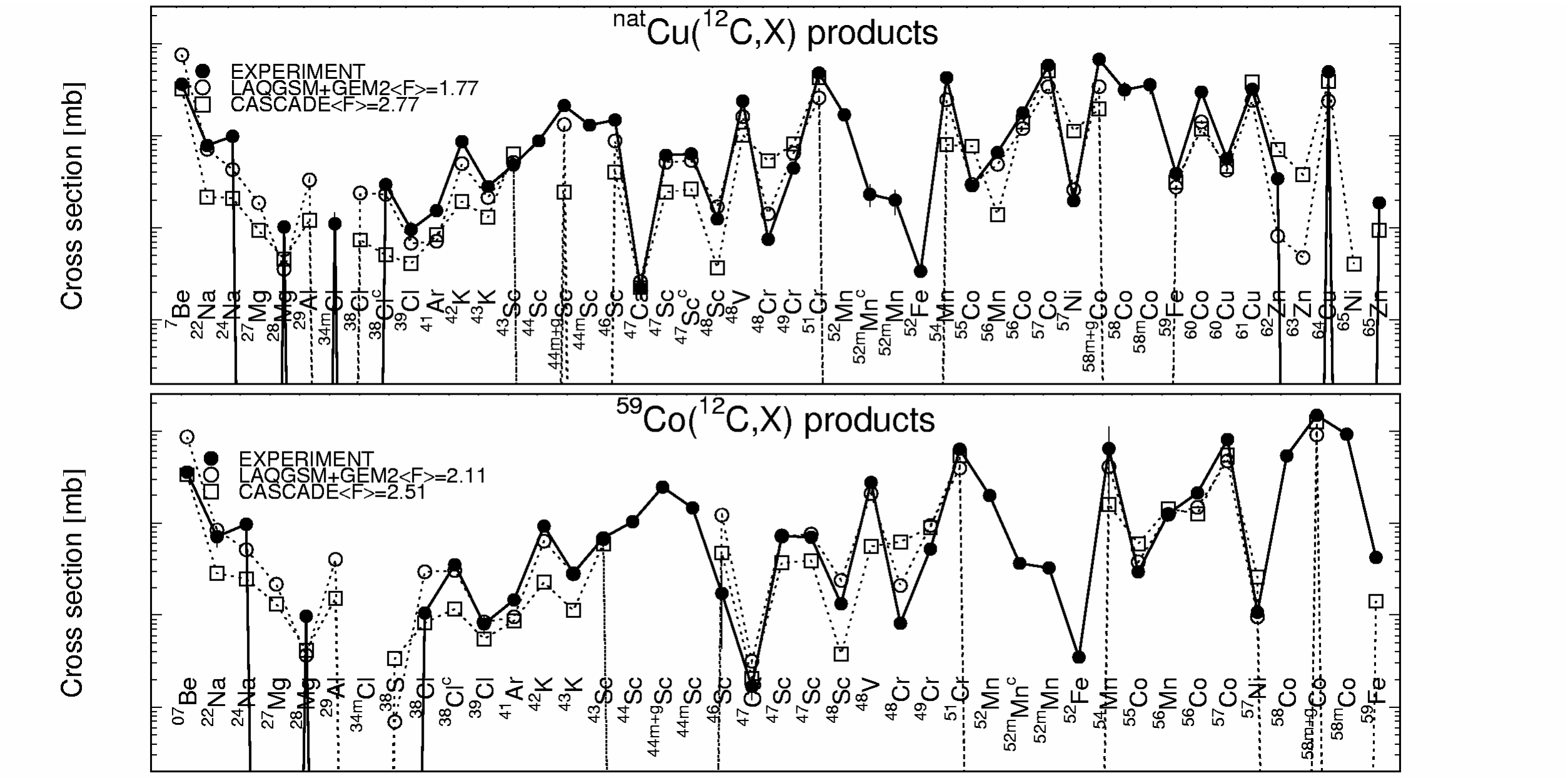

Fig. 3. Nuclide-by-nuclide comparison between the measured and LAQGSM+GEM2 and CASCADE code-simulated SP cross sections in the 0.2 GeV/A $^{12}$C ion irradiation of $^{59}$Co and $^{nat}$Cu.

Table 1

Cross sections (mbarn) for residual product nuclide production in interactions of the 200 MeV/A $^{12}$C ions and the 200 MeV and 2600 MeV protons with thin $^{nat}$Cu sample

| Product | $T_{1/2}$ | Type | $\sigma_{proton}$ $E_p$=200 MeV | $\sigma_{ion}$/ $\sigma_{proton}$ | $\sigma_{proton}$ $E_p$=2600 MeV | $\sigma_{ion}$/ $\sigma_{proton}$ | $\sigma_{ion}$ $E_{12C}$=200 MeV/A |
|---|---|---|---|---|---|---|---|
| $^{65}$Zn | 244.26d | i | 0.888±0.068 | 3.83 | 0.675±0.077 | 5.03 | 3.40±0.49 |
| $^{63}$Zn | 38.47m | i | 2.83±0.25 | - | -- | - | -- |
| $^{62}$Zn | 9.26h | i | 1.73±0.12 | 1.07 | 0.278±0.025 | 6.65 | 1.85±0.19 |
| $^{64}$Cu | 12.700h | i | 21.0±1.5 | 2.34 | 18.6±1.7 | 2.64 | 49.1±5.0 |
| $^{61}$Cu | 3.333h | c | 24.8±2.1 | 1.29 | 9.99±1.05 | 3.19 | 31.9±3.9 |
| $^{60}$Cu | 23.7m | c | 6.81±0.40 | 0.83 | 2.05±0.15 | 2.75 | 5.64±0.58 |
| $^{65}$Ni | 2.51719h | c | -- | - | 0.107±0.011 | - | -- |
| $^{57}$Ni | 35.60h | c | 1.67±0.13 | 1.17 | 0.616±0.052 | 3.18 | 1.96±0.17 |
| $^{60}$Co | 5.2714y | i(m+g) | 12.6±1.0 | 2.37 | 10.2±0.7 | 2.92 | 29.9±3.0 |
| $^{58m}$Co | 9.04h | i(m) | 26.1±1.5 | 1.36 | 14.6±1.2 | 2.43 | 35.6±7.2 |
| $^{58}$Co | 70.86d | i | 13.7±1.0 | 2.28 | 7.33±0.85 | 4.25 | 31.2±7.0 |
| $^{58}$Co | 70.86d | i(m+g) | 39.8±2.1 | 1.68 | 21.9±1.4 | 3.04 | 66.8±4.9 |
| $^{57}$Co | 271.74d | c | 38.1±2.1 | 1.53 | 19.1±1.3 | 3.04 | 58.1±4.3 |
| $^{56}$Co | 77.233d | c | 11.6±0.6 | 1.52 | 5.95±0.41 | 2.97 | 17.7±1.3 |
| $^{55}$Co | 17.53h | c | 1.79±0.11 | 1.59 | 0.964±0.071 | 2.95 | 2.85±0.25 |
| $^{59}$Fe | 44.472d | c | 1.14±0.07 | 3.45 | 1.67±0.12 | 2.34 | 3.91±0.49 |
| $^{52}$Fe | 8.275h | c | -- | -- | 0.142±0.011 | 2.37 | 0.336±0.031 |
| $^{56}$Mn | 2.5789h | c | 2.27±0.11 | 2.87 | 2.83±0.17 | 2.30 | 6.50±0.49 |
| $^{54}$Mn | 312.11d | i | 16.1±0.9 | 2.66 | 15.7±1.0 | 2.72 | 42.7±3.2 |
| $^{52m}$Mn | 21.1m | i(m) | 1.81±0.12 | 1.10 | 1.89±0.14 | 1.05 | 1.98±0.61 |
| $^{52m}$Mn | 21.1m | c | 1.92±0.13 | 1.21 | 2.02±0.15 | 1.15 | 2.31±0.61 |
| $^{52}$Mn | 5.591d | c | 4.90±0.28 | 3.42 | 5.98±0.39 | 2.80 | 16.8±1.3 |
| $^{51}$Cr | 27.7025d | c | 10.7±0.6 | 4.48 | 19.0±1.3 | 2.54 | 48.1±3.7 |
| $^{49}$Cr | 42.3m | c | -- | -- | 2.47±0.19 | 1.79 | 4.41±0.38 |
| $^{48}$Cr | 21.56h | c | -- | -- | 0.325±0.023 | 2.32 | 0.75±0.06 |
| $^{48}$V | 15.9735d | c | 2.13±0.12 | 11.06 | 9.87±0.66 | 2.39 | 23.6±1.7 |
| $^{48}$Sc | 43.67h | i | -- | -- | 0.628±0.039 | 1.98 | 1.24±0.11 |
| $^{47}$Sc | 3.3492d | c | 0.243±0.013 | 25.29 | 3.00±0.18 | 2.05 | 6.37±0.48 |
| $^{47}$Sc | 3.3492d | i | -- | -- | 2.91±0.18 | 2.19 | 6.15±0.46 |
| $^{46}$Sc | 83.79d | i(m+g) | -- | -- | 6.81±0.44 | 2.15 | 14.7±1.1 |
| $^{44m}$Sc | 58.61h | i(m) | 0.244±0.016 | 53.13 | 5.70±0.38 | 2.28 | 13.0±1.0 |
| $^{44}$Sc | 3.97h | i(m+g) | 0.436±0.027 | 48.74 | 9.60±0.62 | 2.21 | 21.3±1.6 |
| $^{44}$Sc | 3.97h | i | 0.210±0.013 | 41.41 | 4.36±0.28 | 2.00 | 8.71±0.66 |
| $^{43}$Sc | 3.891h | c | -- | -- | -- | -- | 4.83±0.54 |
| $^{47}$Ca | 4.536d | c | -- | -- | 0.096±0.012 | 2.27 | 0.219±0.053 |
| $^{43}$K | 22.3h | c | -- | -- | 1.35±0.08 | 2.06 | 2.78±0.21 |
| $^{42}$K | 12.360h | i | -- | -- | 3.90±0.24 | 2.20 | 8.58±0.66 |
| $^{41}$Ar | 109.34m | c | -- | -- | 0.841±0.051 | 1.83 | 1.54±0.14 |
| $^{39}$Cl | 55.6m | c | -- | -- | 0.581±0.041 | 1.63 | 0.947±0.212 |
| $^{38}$Cl | 37.24m | i(m+g) | -- | -- | -- | -- | 2.93±0.39 |
| $^{38}$Cl | 37.24m | c | | | 1.81±0.12 | -- | -- |
| $^{34m}$Cl | 32.00m | i(m) | -- | -- | 0.617±0.048 | 1.78 | 1.10±0.35 |
| $^{38}$S | 170.3m | c | -- | -- | -- | -- | -- |
| $^{29}$Al | 6.56m | c | -- | -- | 1.91±0.35 | -- | -- |
| $^{28}$Mg | 20.915h | c | -- | -- | 0.410±0.025 | 2.50 | 1.02±0.08 |
| $^{27}$Mg | 9.462m | c | | | 1.20±0.12 | -- | -- |
| $^{24}$Na | 14.9590h | c | -- | -- | 3.45±0.22 | 2.85 | 9.84±0.73 |
| $^{22}$Na | 2.6019y | c | -- | -- | -- | -- | 7.78±1.02 |
| $^{7}$Be | 53.29d | i | -- | -- | 8.31±0.68 | 4.34 | 36.0±3.0 |

Table 2

Cross sections (mbarn) for residual product nuclide production in interactions of the 200 MeV/A $^{12}$C ions and the 200 MeV and 2600 MeV protons with thin $^{59}$Co sample

| Product | $T_{1/2}$ | Type | $\sigma_{proton}$ $E_p$=200 MeV | $\sigma_{ion}/\sigma_{proton}$ | $\sigma_{proton}$ $E_p$=2600 MeV | $\sigma_{ion}/\sigma_{proton}$ | $\sigma_{ion}$ $E_{12C}$=200 MeV/A |
|---|---|---|---|---|---|---|---|
| $^{57}$Ni | 35.60h | c | 0.781±0.074 | 1.38 | 0.223±0.020 | 4.84 | 1.08±0.09 |
| $^{58}$Co | 9.04h | i(m) | 41.0 ± 3.8 | 2.24 | 30.1 ±2.7 | 3.05 | 91.9±10.4 |
| $^{58}$Co | 70.86d | i(m+g) | 63.5 ± 5.7 | 2.30 | 47.7 ± 3.9 | 3.06 | 146±11 |
| $^{58}$Co | 70.86d | i | 22.5 ± 2.1 | 2.39 | 17.7 ± 1.8 | 3.03 | 53.7±8.7 |
| $^{57}$Co | 271.79d | c | 49.2 ± 4.4 | 1.64 | 24.2 ± 2.0 | 3.34 | 80.8±6.0 |
| $^{56}$Co | 77.233d | c | 15.8 ± 1.4 | 1.33 | 5.63 ±0.45 | 3.73 | 21.0±1.6 |
| $^{55}$Co | 17.53h | c | 2.53 ± 0.23 | 1.17 | 0.762±0.065 | 3.87 | 2.95±0.24 |
| $^{59}$Fe | 44.472d | c | -- | | 0.537±0.048 | 7.80 | 4.19±0.48 |
| $^{52}$Fe | 8.275h | c | 0.255±0.024 | 1.36 | 0.120±0.011 | 2.89 | 0.347±0.031 |
| $^{56}$Mn | 2.5789h | c | 4.54 ±0.41 | 2.75 | 4.99 ±0.41 | 2.51 | 12.5±0.9 |
| $^{54}$Mn | 312.11d | i | 35.8 ±3.2 | 1.79 | 21.3 ± 1.8 | 3.00 | 64.0±4.8 |
| $^{52m}$Mn | 21.1m | i(m) | 4.57 ±0.43 | 0.72 | 2.45 ±0.22 | 1.33 | 3.27±0.49 |
| $^{52m}$Mn | 21.1m | c | 4.82 ±0.45 | 0.76 | 2.61 ±0.23 | 1.40 | 3.66±0.49 |
| $^{52}$Mn | 5.591d | c | 12.0 ±1.1 | 1.66 | 6.14 ±0.50 | 3.24 | 19.9±1.5 |
| $^{51}$Cr | 27.7025d | c | 31.4 ±2.9 | 2.00 | 21.4 ±1.8 | 2.93 | 62.8±4.8 |
| $^{49}$Cr | 42.3m | c | 2.82 ±0.28 | 1.83 | 2.61 ±0.24 | 1.98 | 5.17±0.42 |
| $^{48}$Cr | 21.56h | c | 0.252±0.023 | 3.23 | 0.317±0.027 | 2.57 | 0.814±0.062 |
| $^{48}$V | 15.9735d | c | 8.43 ±0.75 | 3.26 | 10.6 ± 0.9 | 2.59 | 27.5±2.0 |
| $^{48}$Sc | 43.67h | i | 0.186±0.017 | 7.15 | 0.625±0.051 | 2.13 | 1.33±0.10 |
| $^{47}$Sc | 3.3492d | c | 1.14 ±0.10 | 6.16 | 3.20 ±0.27 | 2.19 | 7.02±0.53 |
| $^{47}$Sc | 3.3492d | i | 1.09 ±0.10 | 6.60 | 3.12 ±0.26 | 2.30 | 7.19±0.54 |
| $^{46}$Sc | 83.79d | i(m+g) | 2.52 ±0.26 | 6.87 | 7.42 ±0.60 | 2.33 | 17.3±1.3 |
| $^{44m}$Sc | 58.61h | i(m) | 1.27 ±0.12 | 11.42 | 5.84 ±0.48 | 2.48 | 14.5±1.1 |
| $^{44}$Sc | 3.97h | i(m+g) | 2.41 ±0.22 | 10.12 | 10.9 ± 0.9 | 2.24 | 24.4±1.8 |
| $^{44}$Sc | 3.97h | i | 1.17 ±0.11 | 8.80 | 5.29 ±0.44 | 1.95 | 10.3±0.8 |
| $^{43}$Sc | 3.891h | c | 0.491±0.048 | 14.01 | 3.46 ±0.31 | 1.99 | 6.88±0.59 |
| $^{47}$Ca | 4.536d | c | 0.051±0.009 | 3.25 | 0.082±0.010 | 2.02 | 0.166±0.047 |
| $^{43}$K | 22.3h | c | 0.107±0.010 | 26.73 | 1.42 ±0.11 | 2.01 | 2.86±0.21 |
| $^{42}$K | 12.360h | i | 0.357±0.033 | 25.77 | 4.17 ±0.35 | 2.21 | 9.20±0.71 |
| $^{41}$Ar | 109.34m | c | 0.036±0.004 | 40.83 | 0.836±0.070 | 1.76 | 1.47±0.16 |
| $^{39}$Cl | 55.6m | c | -- | | 0.566±0.049 | 1.40 | 0.795±0.184 |
| $^{38}$Cl | 37.24m | i(m+g) | -- | | 1.93 ±0.17 | 1.83 | 3.54±0.39 |
| $^{38}$Cl | 37.24m | c | -- | | 2.00 ±0.17 | 0.53 | 1.06±0.19 |
| $^{34m}$Cl | 32.00m | i(m) | -- | | 0.704±0.061 | | -- |
| $^{38}$S | 170.3m | c | -- | | 0.066±0.007 | | -- |
| $^{29}$Al | 6.56m | c | -- | | 2.56 ±0.24 | | -- |
| $^{28}$Mg | 20.915h | c | -- | | 0.432±0.035 | 2.27 | 0.979±0.074 |
| $^{27}$Mg | 9.462m | c | -- | | 1.53 ±0.14 | | -- |
| $^{24}$Na | 14.9590h | c | -- | | 3.77 ±0.31 | 2.55 | 9.62±0.72 |
| $^{22}$Na | 2.6019y | c | -- | | 2.46 ±0.22 | 2.87 | 7.05±1.62 |
| $^{7}$Be | 53.29d | i | -- | | 8.78 ±0.89 | 4.07 | 35.7±2.9 |

Table 3

Cross sections (mbarn) for residual product nuclide production in interactions of the 200 MeV/A $^{12}$C ions and the 200 MeV and 2600 MeV protons with thin $^{27}$Al sample

| Product | $T_{1/2}$ | Type | $\sigma_{proton}$ $E_p$=200 MeV | $\sigma_{ion}/\sigma_{proton}$ | $\sigma_{proton}$ $E_p$=2600 MeV | $\sigma_{ion}/\sigma_{proton}$ | $\sigma_{ion}$ $E_{12C}$=200 MeV/A |
|---|---|---|---|---|---|---|---|
| $^{24}$Na | 14.9590h | c | 9.8±0.7 | 3.51 | 10.6±0.9 | 3.25 | 34.4±2.8 |
| $^{22}$Na | 2.6019y | c | 15.1±0.9 | 2.42 | 11.7±0.9 | 3.12 | 36.5±3.6 |
| $^{7}$Be | 53.29d | i | 1.48±0.10 | 23.2 | 9.2±0.7 | 3.74 | 34.4±2.8 |

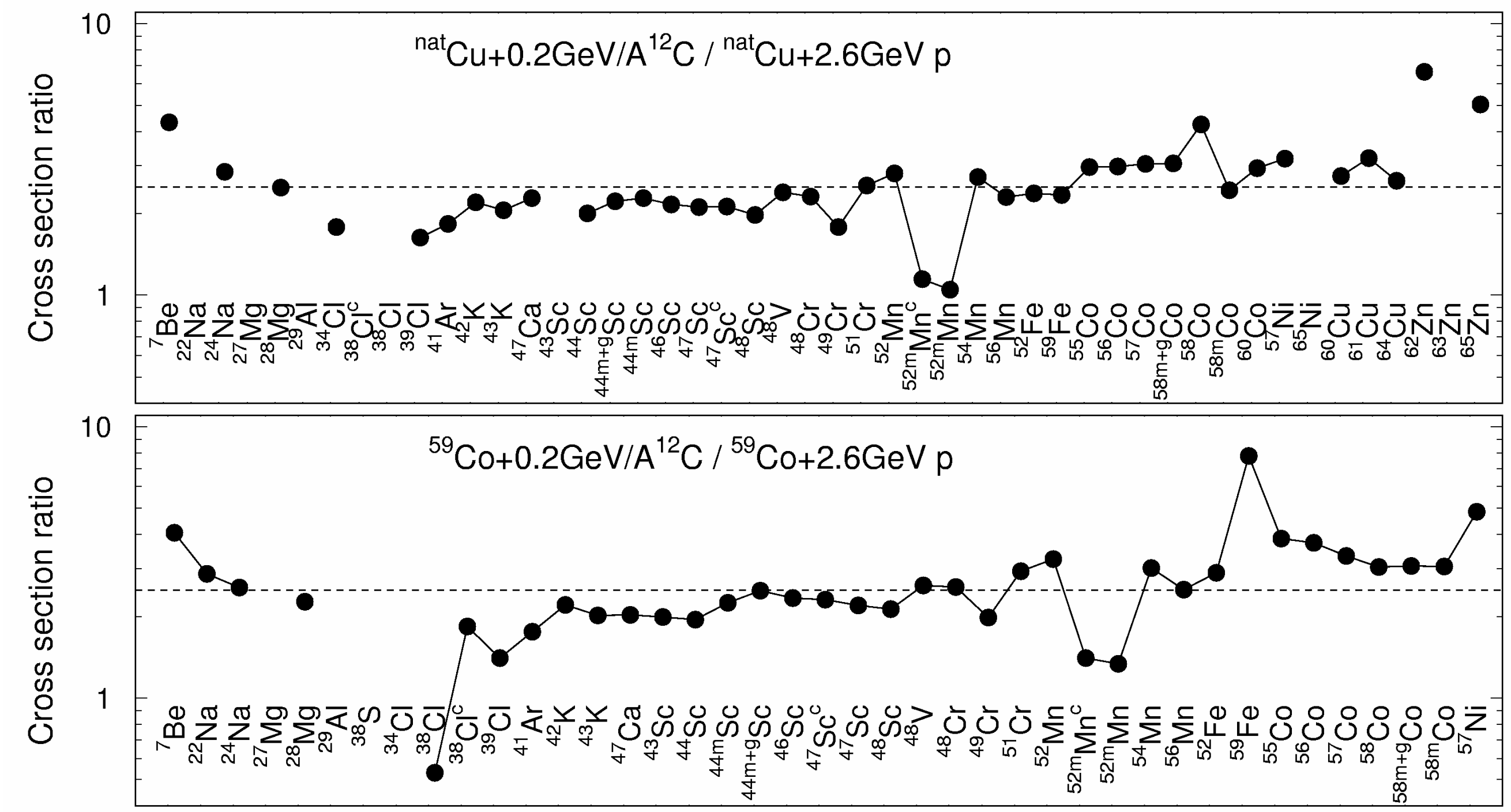

Fig. 4. The ratios of the SP cross sections in the 0.2 GeV/A $^{12}$C ion irradiation of $^{59}$Co and $^{nat}$Cu to the SP cross section in the 0.2 GeV proton irradiation of the same nuclides.

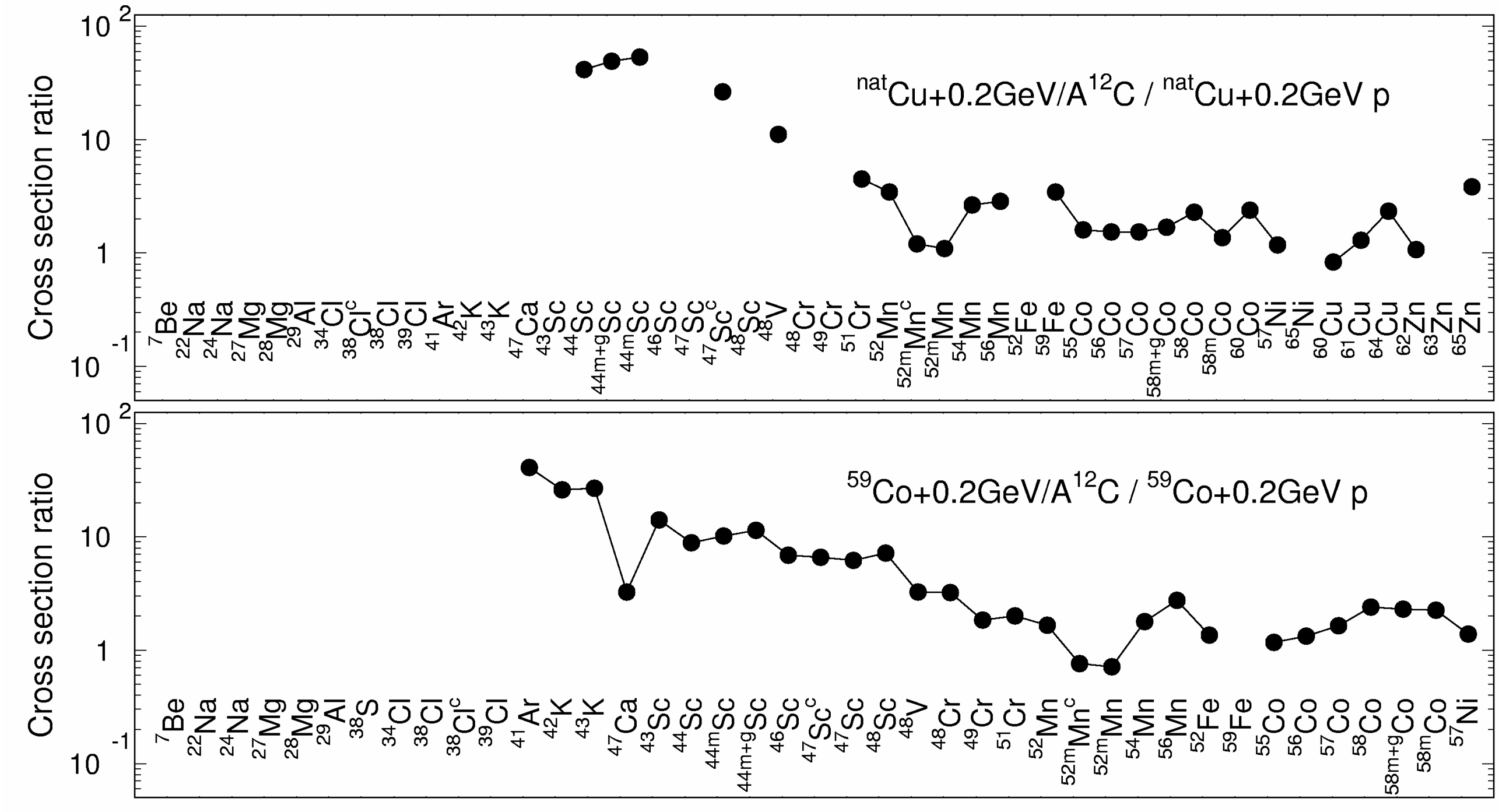

Fig. 5. The ratios of the SP cross sections in the 0.2 GeV/A $^{12}C$ ion irradiation of $^{59}Co$ and $^{nat}Cu$ to the SP cross section in the 2.6 GeV proton irradiation of the same nuclides.

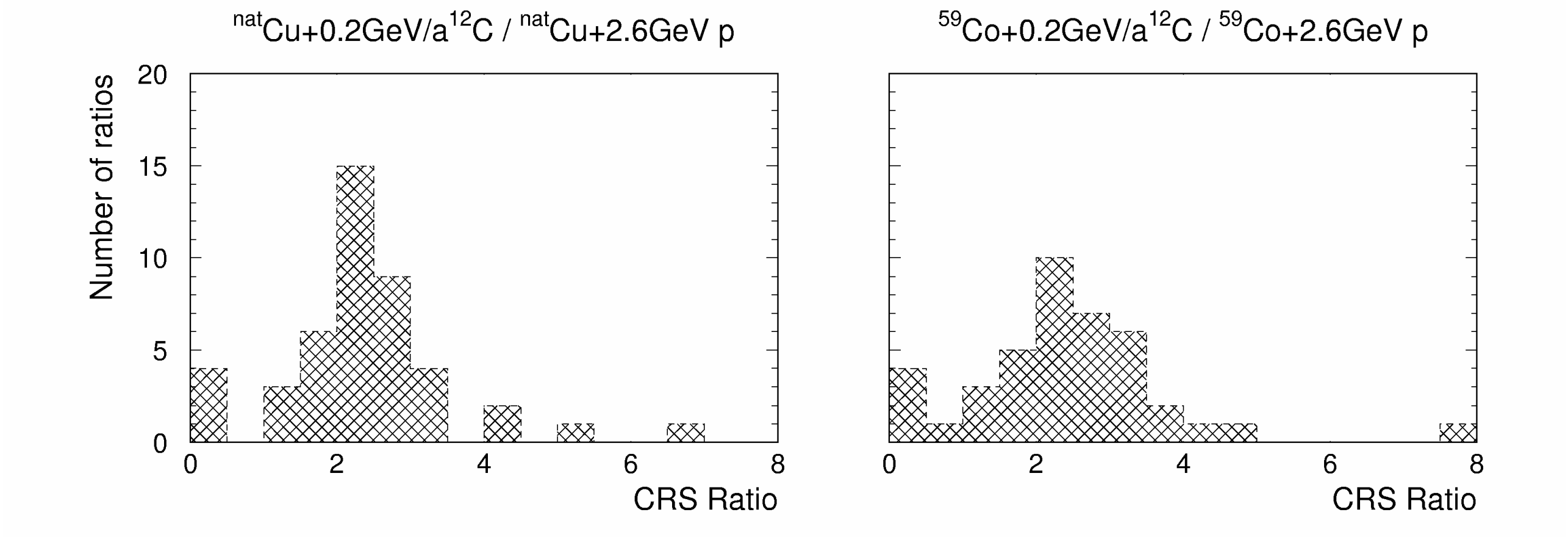

Fig. 6. Statistics of the ratios of the SP cross sections in the 0.2 GeV/A $^{12}C$ ion irradiation of $^{59}Co$ and $^{nat}Cu$ to the SP cross section in the 2.6 GeV proton irradiation of the same nuclides.